# Time as a parameter of statistical ensemble

Sergei Viznyuk

**Abstract**

The notion of time is derived as a parameter of statistical ensemble representing the underlying system. Varying population numbers of microstates in statistical ensemble result in different expectation values corresponding to different *times*. We show a single parameter which equates to the notion of time is logarithm of the total number of microstates in statistical ensemble. We discuss the implications of proposed model for some topics of modern physics: Poincaré recurrence theorem vs. Second Law of Thermodynamics, matter vs. anti-matter asymmetry of the universe, expansion of the universe, Big Bang.

> The idea of any (thermodynamic) system always approaching a "state" (equilibrium) in which it behaves as if it occupied an immense multitude of microstates at the same time (ensemble), even if it started from a perfectly concrete microstate, is still quite puzzling.
>
> --Gemmer J., et al [1]

## 1. INTRODUCTION

Efforts to understand the nature of time span centuries [2]. Some concepts, ordered roughly by the time of their appearance are:

- Time is an absolute entity, an objective reality, existing and flowing independently of the observed system [3].
- Time is the forth dimension of a spacetime continuum, called Minkowski space [4].
- Time is a redundant notion which, given sufficient effort, can be eliminated from all equations of physics in favor of more tangible quantities, like distance, energy, etc. [5,6].
- Time is an ordering parameter for probabilities of observable states of the system [7-9].

We derive the notion of time using basic principle of statistical mechanics, that any state of the system can be described as a combination of microscopic configurations (microstates). Such combination is called statistical ensemble. Varying population numbers of microstates in statistical ensemble result in different expectation values corresponding to different *times*.

We show a single parameter which equates to the notion of time is logarithm of the total number of microstates in statistical ensemble. The presented model of time:
- resolves contradiction between Poincaré recurrence theorem and Second Law of Thermodynamics
- offers explanation for matter vs. anti-matter asymmetry of the universe
- predicts accelerating expansion of the universe
- demonstrates the early time sequence of Big Bang is similar to the process of *spontaneous emission*

## 2. STATISTICAL ENSEMBLE

In this section we provide definitions for the notions used in the model. We use the concept of *microstate* as base construct.

- $G$ – set of all accessible and distinct microstates $|i\rangle$ which can be distinguished through an observable $X$ (1)
- $M$ – cardinality of set $G$ (2)
- *statistical ensemble* – sampling with replacement of $N$ microstates from set $G$ (3)
- *equlibrium*[1] – statistical ensemble with equal population $n_i$ of microstates (4)

The probability of sampling a specific microstate from statistical ensemble in *equilibrium* is $p_{eq} = 1/M$ for all microstates $|i\rangle \in G$. As defined, the statistical ensemble has multinomial probability distribution centered around equilibrium:

$$P(\{n_i\}; N, p_{eq}) = \frac{N!}{\prod_{i \in G}(n_i)!} (p_{eq})^N = \Omega_N \cdot (p_{eq})^N = e^{S_N + N \ln p_{eq}} \quad (5)$$

, where $S_N$ is the entropy of statistical ensemble measured in *nats*:

$$S_N = \ln(\Omega_N) \quad \text{, and } \Omega_N \text{ is the } statistical\ weight;\quad \Omega_N = \frac{N!}{\prod_{i \in G}(n_i)!} \quad (6)$$

The unit entropy[2], in nats per microstate is

$$s_N = \frac{S_N}{N} = \frac{\ln(\Omega_N)}{N} = \frac{1}{N}\left[\ln \Gamma(N+1) - \sum_{i \in G} \ln \Gamma(n_i + 1)\right] \quad (7)$$

, where $\Gamma(x)$ is *gamma* function. Using *Stirling approximation* $\ln(n!) \cong n \cdot ln(n) - n$, when $N \gg 1$, $n_i \gg 1$ formula (7) for unit entropy can be rewritten into commonly used expression[3]:

$$s_N = -\sum_{i \in G} \frac{n_i}{N} \ln \frac{n_i}{N} = -\sum_{i \in G} p_i \ln p_i \quad (8)$$

## 3. TIME MODEL

The expectation value $X$ is determined by population numbers of microstates[4] from set $G$:

$$X = F(\{p_i\}) \quad \text{, where} \quad p_i = \frac{n_i}{N} \quad (9)$$

From (9):

$$\partial X = \sum_{i \in G} \frac{\partial F}{\partial p_i} \cdot \left(\frac{\partial n_i}{N} - \frac{n_i}{N} \cdot \frac{\partial N}{N}\right) = \sum_{i \in G} \frac{\partial F}{\partial p_i} \cdot (q_i - p_i) \cdot \partial \ln N \quad (10)$$

, where $\{q_i\}$ is the probability distribution of variations $\{\Delta n_i\}$ from the current state $\{n_i\}$, i.e. $\Delta n_i = q_i \cdot \Delta N$. In (10) we assumed $n_i \gg \Delta n_i \gg 1$; $N \gg \Delta N \gg 1$ in order to use differential math. We only use this assumption in order to extract from (10) the definition of time.

We rewrite (10) as

$$\partial X = \sum_{i \in G} \frac{\partial F}{\partial p_i} \cdot (q_i - p_i) \cdot \partial t \quad \text{, where} \quad t = \ln N \quad (11)$$

---

[1] As defined, the equilibrium here is *local* with respect to observable $X$
[2] The unit entropy here is the *microstate entropy* [10], as opposed to *thermodynamic entropy* defined for ensemble of observable entities
[3] In practical situations there is hardly any advantage of using approximate formula (8) over exact (7), since computing $\ln \Gamma(x)$ is about as easy as computing $\ln x$.
[4] Expression (9) for the expectation value is a generalization over $X = Tr(\rho X)$, where $\rho$ is the density matrix representing microstate distribution, and $X$ is the operator matrix of observable $X$

Entities in (1-11) have been defined within the context of a given observable $X$. To indicate that, we shall append subscript $X$ where necessary when providing the following definition of time:

- the *proper time*[5] of observable $X$ is $\quad t_X = \ln N_X \quad$, where $\quad N_X = \sum_{i \in G_X} n_i \quad$ (12)

We shall omit subscript $X$ from entity names, except where necessary to distinguish between different observables. To define the direction of time change $\Delta t$ we note that arbitrary variation $\Delta N = \sum \Delta n_i$ can be decomposed into two parts:

$$\Delta N = \sum_{\Delta n_i > 0} \Delta n_i + \sum_{\Delta n_i < 0} \Delta n_i = \Delta N|_{\Delta t > 0} + \Delta N|_{\Delta t < 0} \quad , \text{ where we define:}$$

the positive $\Delta t > 0$ direction of time change is when $\quad \Delta N = \Delta N|_{\Delta t > 0} \equiv \sum_{\Delta n_i > 0} \Delta n_i \quad$ (13)

, and negative $\Delta t < 0$ direction of time change is when $\quad \Delta N = \Delta N|_{\Delta t < 0} \equiv \sum_{\Delta n_i < 0} \Delta n_i \quad$ (14)

A case with $\Delta N = 0$ is equivalent to the switch to a different *current* state $\{n_i\}$ if $\Delta n_i \neq 0$ for some microstates.

The probability of sampling a specific microstate from set $G$ is independent of the microstate. Therefore we expect the probabilities $q_i$ of the increments $\{\Delta n_i\}$ in the positive direction of time to be $\quad q_i(\{n_i\}) \equiv p_{eq} \quad \forall \quad |i\rangle \in G \quad$ (15)

We consider the current state $\{n_i\}$ has progressed from some state in the *past* through positive increments $\{\Delta n_i > 0\}$ having probabilities (15). Stepping back in time means decrementing from current state $\{n_i\}$ by $\{\Delta n_i < 0\}$ with the same probabilities (15), subject to condition $n_i > 0$.

For a simple case when operator matrix of observable $X$ is diagonal in microstate representation we have:

$$X = F(\{p_i\}) \equiv \sum_{i \in G} X_{ii} \cdot p_i \quad (16)$$

From (11, 15, and 16) we then obtain the exponential decay[6] law for observable $X$:

$$\frac{\partial X}{\partial t} = \sum_{i \in G} \frac{\partial F}{\partial p_i} \cdot (q_i - p_i) = \sum_{i \in G} X_{ii} \cdot (p_{eq} - p_i) = X_{eq} - X \quad (17)$$

$$X = X_{eq} + (X_0 - X_{eq}) \cdot exp(t_0 - t) \quad ; \quad t \geq t_0 \quad (18)$$

Here $X_0$ is the expectation value at time $t = t_0$, and $X_{eq}$ is the expectation value in equilibrium. From (12,15) we obtain the relation between *proper time* scales of different observables $X$ and $Y$. We append subscripts $X$ and $Y$ to distinguish between entities belonging to different observables:

$$\frac{\Delta t_X}{\Delta t_Y} = \frac{\Delta N_X}{\Delta N_Y} \cdot \frac{N_Y}{N_X} = \frac{M_X}{M_Y} \cdot \frac{N_Y}{N_X} = \frac{n_Y}{n_X} \quad (19)$$

---

[5] Perhaps the alternative name for the *proper time of observable X* could be *the internal clocks of observable X*

[6] The differential equations here are valid in approximation $n_i \gg \Delta n_i \gg 1$; $N \gg \Delta N \gg 1$. If we step back in time, this condition will inevitably violate, as we demonstrate in numeric analysis section. Therefore in general the decay law (18) is only valid for $t \geq t_0$

, where $n_X = N_X/M_X$; $n_Y = N_Y/M_Y$ are the average microstate population numbers from sets $G_X$ and $G_Y$. Relation (19) needs interpretation. Suppose the observable $Y$ includes the observer with the wall clocks. Suppose the observable $X$ is some other observed entity, for example an atom or even a macroscopic object. Then, $\Delta t_Y$ is the time interval shown by the wall clocks, and $\Delta t_X$ is the time inverval by *internal clocks* of observable $X$. The relation (19) shows that the *internal clocks* of observable $X$ run the faster relative to wall clocks the smaller is the average population of microstates from set $G_X$. We may conclude that all transient phenomena are exhibited by the observables with relatively low population numbers of microstates from their respective sets. Impirically this is an obvious result, since it takes a smaller increment $\Delta N$ to equalize the set $\{n_i\}$ of small unequal numbers, than a set of large proportionally unequal numbers.

## 4. NUMERIC ANALYSIS

In this section we investigate the temporal evolution of an observable $X$ represented by randomly chosen Hermitian operator matrix $X$, along with temporal evolution of probabilities $\{p_i\}$, and of unit entropy (7) according to the presented time model. For the numeric calculations we follow linear algebra model:

$$X = F(\{p_i\}) \equiv \text{Tr}(\rho X) \qquad (20)$$

, where $X$ is the operator matrix of observable $X$; and $\rho$ is the density matrix. The population numbers $\{n_i\}$; $|i\rangle \in G$ provide complete description of underlying system within the context of observable $X$. Therefore in observation space the statistical ensemble is a *state vector*[7]:

$$|n\rangle = \sum_{i \in G} C_i \cdot |i\rangle \qquad , \text{where} \qquad C_i^* \cdot C_i = n_i \qquad (21)$$

From (21):
$$C_i = \sqrt{n_i} \cdot exp(i \cdot \varphi_{n_i}) \qquad (22)$$

And the density matrix is
$$\rho_{ij} = C_i^* \cdot C_j = \frac{\sqrt{n_i n_j}}{N} \cdot exp\left[i \cdot \left(\varphi_{n_j} - \varphi_{n_i}\right)\right] \qquad (23)$$

The expectation value (20) should not change when all population numbers $n_i$ are multiplied by the same positive number. Therefore $\varphi_{n_i}$ has to be of the form

$$\varphi_{n_i} = \omega \cdot \ln n_i \quad , \text{where } \omega \text{ is not dependent on } \{n_i\} \qquad (24)$$

From (24) and (23) the density matrix[8] is
$$\rho_{ij} = \frac{\sqrt{n_i n_j}}{N} \cdot exp\left[i \cdot \omega \cdot \ln \frac{n_j}{n_i}\right] \qquad (25)$$

Figures 1-7 show temporal evolution of probabilities $\{p_i\}$, expectation value (20), and microstate entropy (7). The initial state $\{n_i\}$ has been randomly generated for proper time $t_0 = \ln(N = 10000) = 9.21$. The evolution in positive direction of time has been calculated in steps of $\Delta N = 1$ with probability (15) for a microstate to increment its population by 1 in one step. The *devolution* in negative direction of time has been calculated in steps of $\Delta N = -1$ with the same probability (15) for a microstate to decrement its population by 1 in one step, subject to

---

[7] The classical thermodynamic ensembles represent multiple copies of the same system, e.g. a gas of uncorrelated identical particles. The statistical ensemble defined in this article represents a single copy of underlying system. As a consequence the thermodynamic entropy $S_T = -\text{Tr}(\rho \cdot \ln \rho) \equiv 0$, where $\rho$ is density matrix (25)

[8] All entities here, including $\omega$, and density matrix $\rho$ are defined within the context of observable $X$, and generally should be carrying subscript $X$

condition $n_i > 0$. The devolution in negative direction of time is shown in different line color, with line originating at the same time $t_0$ as the evolution in positive direction of time. The expectation value $X$ has been calculated using (20) with randomly generated Hermitian operator matrix $X$, and with density matrix (25). While horizontal axes are logarithmic by $N$, they are linear by proper time $t = \ln N$. The Matlab code used in calculations and to generate the graphs is available at http://phystech.com/download/ensemble_dynamics.m

Figures 1-7 demonstrate time evolution from the initial state $\{n_i\}$ at $t_0$ toward equilibrium as $t \to \infty$. The presented graphs also show the *most likely history* of the system for $t < t_0$. If we step back in time from the current state, according to the presented time model, by randomly decrementing population numbers $\{n_i\}$, we shall arrive to a point in time $t_B$ with statistical ensemble only having one populated microstate. In time interval between $t = \ln(N = 1) = 0$, and $t = t_B$ the system is represented by one populated microstate (Figures 1, 5), and has entropy $S_N = 0$ (Figures 4, 7).

Figures 2, 3, and 6 demonstrate time dynamics of expectation value $X$ for three different parameter settings. Figures 2, 3 show what could be interpreted as transition from excited to equilibrium state in two-level system, for two values of frequency[9] $\omega$. Figures 6, 7 demonstrate transition in three-level system. The temporal evolution of expectation value in presented model resembles *spontaneous emission*. The system stays in excited state between times $t = 0$, and $t = t_B$. Time $t = t_B$ marks the beginning of transition to equilibrium. For three-level system on Figures 6, 7 the transition happens in two stages, with system pausing between first and second transitions in intermediate quasi-stable state.

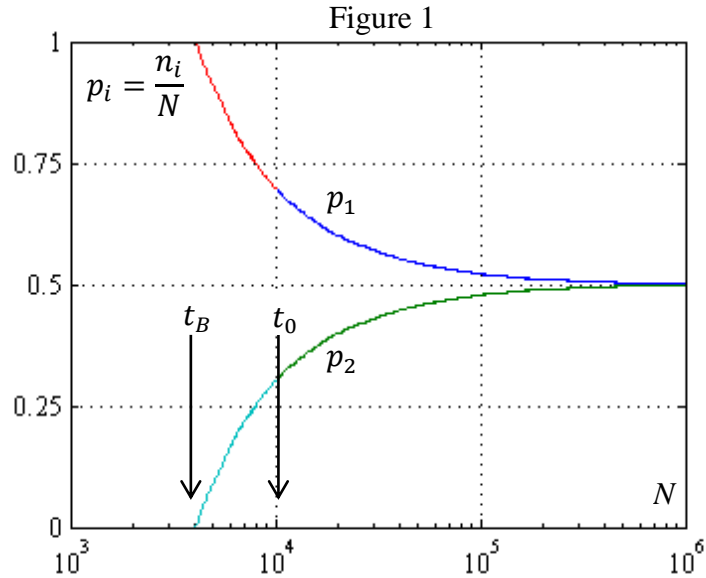

Time dynamics of microstate populations for ensemble of microstates with cardinality $M=2$. The initial state has been randomly generated for $N = 10^4$, corresponding to proper time $t_0 = \ln(10^4) = 9.21$

---

[9] Oscillations on Figures 3, 6 resemble *Rabi oscillation* of expectation value. Therefore we may draw a parallel between parameter $\omega$ in presented model and *Rabi frequency*

Figure 2

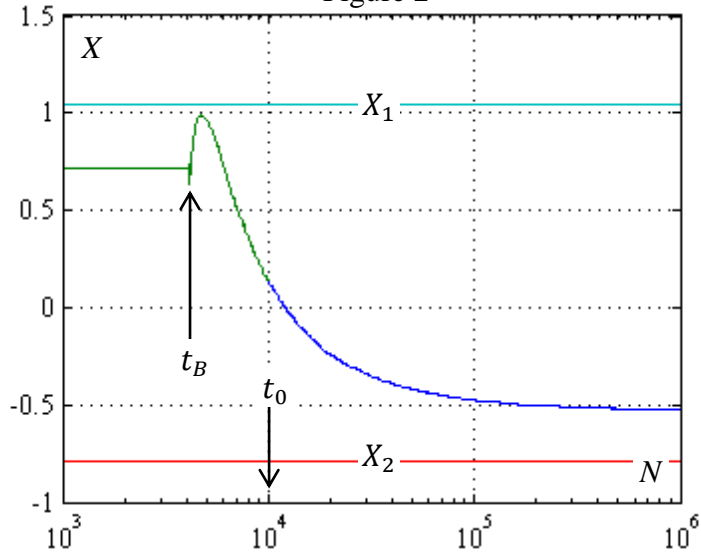

Time dynamics of expectation value $X$ corresponding to time evolution of population numbers on Figure 1, for ensemble of microstates with cardinality $M=2$. Two horizontal lines are eigenvalues $X_1$ and $X_2$ of operator matrix $X$. The graph for $X$ has been calculated using formula (20) with density matrix (25) where $\omega = 1$

Figure 3

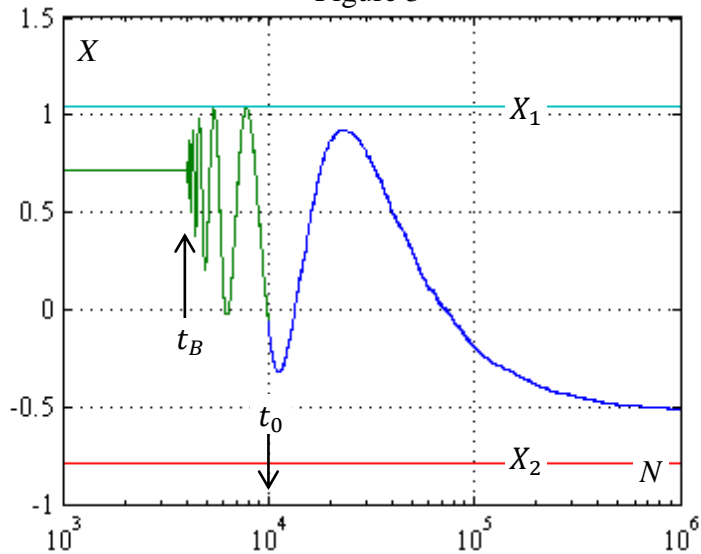

Time dynamics of expectation value $X$ corresponding to time evolution of population numbers on Figure 1, for ensemble of microstates with cardinality $M=2$. Two horizontal lines are eigenvalues $X_1$ and $X_2$ of operator matrix $X$. The graph for $X$ has been calculated using formula (20) with density matrix (25) where $\omega = 8$

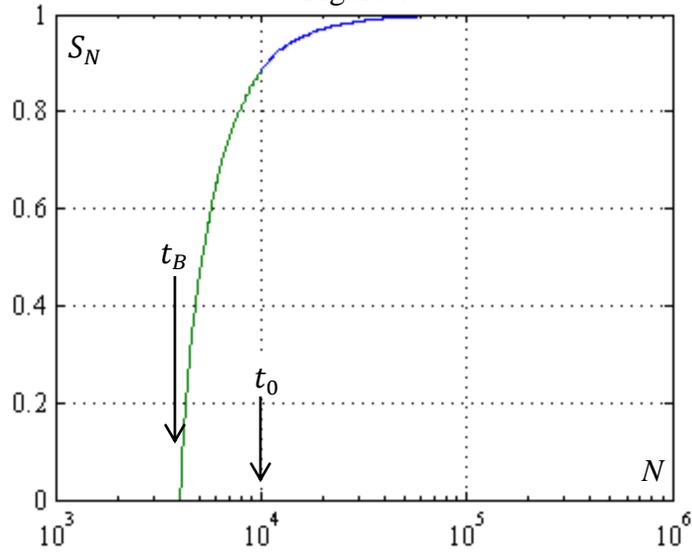

Time dynamics of microstate entropy $s_N$ calculated using formula (7), corresponding to time evolution of population numbers on Figure 1 for ensemble of microstates with cardinality $M=2$. If magnified, the entropy curve shows stochastic behavior, with small-scale entropy fluctuations

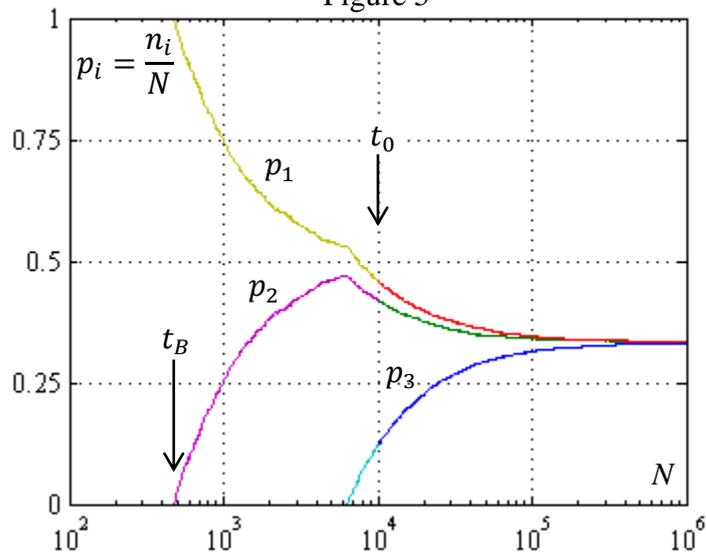

Time dynamics of microstate populations for ensemble of microstates with cardinality $M=3$. The initial state has been randomly generated for $N = 10^4$, corresponding to proper time $t_0 = \ln(10^4) = 9.21$

### Figure 6

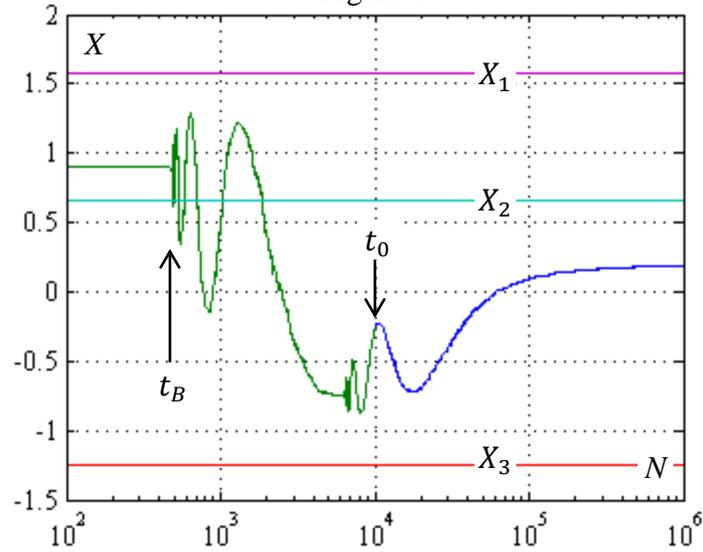

Time dynamics of expectation value $X$ corresponding to time evolution of population numbers on Figure 5, for ensemble of microstates with cardinality $M=3$. Three horizontal lines are eigenvalues $X_1$, $X_2$, and $X_3$ of operator matrix $X$. The graph for $X$ has been calculated using formula (20) with density matrix (25) where $\omega = 5$

### Figure 7

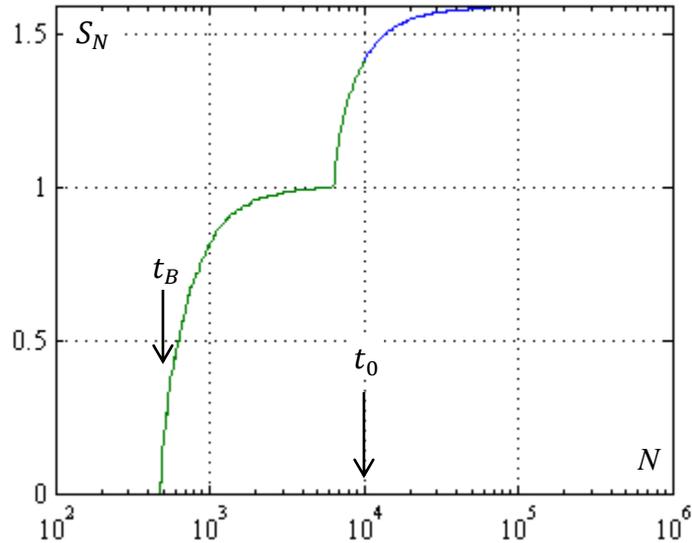

Time dynamics of microstate entropy $s_N$ calculated using formula (7), corresponding to time evolution of population numbers on Figure 5 for ensemble of microstates with cardinality $M=3$. If magnified, the entropy curve shows stochastic behavior, with small-scale entropy fluctuations

## 5. DISCUSSION

The presented time model has the following properties:
1. Time is defined as the *proper time* of a given observable. The expectation value of the observable is determined by population numbers of microstates in statistical ensemble representing the underlying system. The notion of time is not defined for microstate.
2. Time has a beginning at $t = 0$ corresponding to a statistical ensemble with total population $N = 1$
3. Time is quantized with time quantum equal $\Delta t = \ln(N + 1) - \ln(N) \cong 1/N$. Time can be approximated by continuous variable $t$ when considered time scales $\gg 1/N$. The time quantum is decreasing with time as $1/N = exp(-t)$.
4. The evolution in time is *time-reversible stochastic process*.
5. The *measurement* of the observable is the *sampling* of particular combinations of microstates from statistical ensemble. Such combinations of microstates are the *eigenvectors* of the observable in microstate representation.

Some of the implications of the presented time model are:
1. The statistical ensemble representing the underlying system grows with each time increment. Therefore the condition of confinement in Poincaré recurrence theorem [11] is not satisfied. It is straightforward to prove the total entropy (6) of statistical ensemble always increases in positive direction of time, and always decreases or stays equal to zero in negative direction of time. The same is true *in average*[10] for the unit entropy (7).
2. Consider the observable $X$ to be the universe. Using suitable transformation one can build ensemble representation in terms of *matter* and *anti-matter* eigenstates instead of generic microstates. Assuming matter and anti-matter eigenstates differ only through some symmetry transformation, and given current non-equilibrium state of the universe, we expect the population numbers of matter and anti-matter eigenstates to be different. The presented time model predicts the population numbers of anti-matter eigenstates will equalize with those of matter, as the universe approaches equilibrium.
3. From general consideration we expect the volume of the universe to be proportional to the number of microstates $N$ in statistical ensemble representing the universe. Therefore the volume of the universe should grow as $N = exp(t)$, where $t$ is the universe proper time. Thus the presented model predicts *accelerating expansion* of the universe.
4. Stepping back in time from the current non-equilibrium state of the universe inevitably leads to a point in time $t_B$, at which the statistical ensemble is reduced to a single populated microstate, as shown on Figures 1, 5. The entropy $s_N = 0 \ \forall \ 0 \leq t \leq t_B$ as shown on Figures 4, 7. To the observer who is part of the current non-equilibrium state the early time sequence of the universe *appears* as the process of *spontaneous emission*:
   A. The universe is in zero-entropy state during $0 \leq t \leq t_B$
   B. At time $t = t_B$ the universe starts decaying from zero-entropy state to equilibrium, as demonstrated on Figures 2, 3, 6.

---

[10] The unit entropy (7) experiences small fluctuations in opposite direction from mean path due to stochastic nature of time evolution process, as evident if we magnify small fragments of the entropy curves on Figures 4, 7

We can evaluate $t_B$ by noting that decrements $\{\Delta n_i\}$ have multinomial probability distribution (5) with variance $\sigma^2 = \Delta N \cdot p_{eq} \cdot (1 - p_{eq})$. At time $t = t_B$ there remains only one populated microstate with population variance

$$\sigma_B^2 = (N - N_B) \cdot p_{eq} \cdot (1 - p_{eq}) \cong (\Delta N_B)^2 \qquad (26)$$

, where $N_B$ is the total population of statistical ensemble at $t = t_B$; $N_B$ is roughly equal to the difference in population of microstates with maximum population level and the next biggest one at *current* time *t*:

$$N_B \cong n_{i,max} - n_{j,next\ biggest} \qquad (27)$$

With $N_B$, and $\Delta N_B$ given by (27) and (26):

$$t_B = \ln N_B \pm \frac{\Delta N_B}{N_B} \qquad (28)$$